\documentclass[preprint,12pt]{elsarticle}
\usepackage{amsmath,amsfonts,amssymb}
\usepackage{pdfsync}
\usepackage{ifpdf}
\ifpdf
\usepackage{hyperref}
\else
\usepackage[hypertex]{hyperref}
\fi
\usepackage{graphicx}

\biboptions{sort&compress}
\newcommand{\f}{\frac}

\newcommand{\m}{\mathbf}
 \newcommand*{\OrigAA}{}
\let\OrigAA\AA

\renewcommand*{\AA}{%
  {\fontfamily{ptm}%
  \selectfont%
  \OrigAA%
  \selectfont}%
}
\journal{Journal of Molecular Liquids}
\begin{document}
\begin{frontmatter}
\title{New version of the fluctuation
Hamiltonian for liquids near the critical point}
\author{V.L. Kulinskii}
\ead{kulinskij@onu.edu.ua}%

\author{N.P. Malomuzh}
\ead{mnp@normaplus.com}%
\address{Department of
Theoretical Physics, Odessa National University, Dvoryanskaya
2, 65026 Odessa, Ukraine}

%\date{\today}
\begin{abstract}
We propose new canonical form of the
fluctuational Hamiltonian which takes
into account the fact that in the
vicinity of the critical point there
are two fluctuating fields. They are
the field of the number density and the entropy.
The proposed canonical form is based on the
$D_5$ catastrophe.  In contrast to the standard
approach of Landau-Ginzburg Hamiltonian which
is based on $A_3$ catastrophe the canonical
form proposed for the fluctuational Hamiltonian
naturally includes the asymmetric coupling
between the fields.
\end{abstract}
%\pacs{05.70.Jk, 64.60.Fr, 64.70.Fx}
\end{frontmatter}

\section{Introduction}\label{sec_intro}
The modern theory of the critical phenomena
in liquids is based on the Landau-Ginsburg
fluctuation Hamiltonian:
\begin{equation}
\label{lg_ham}
H_{LG}[\psi  (\m{r})] =
\int {d\m{r}{\left[ { \frac{1}{2}(\nabla \psi(\m{r})
)^{2}  - h_{\psi}\,\psi(\m{r})  +
{\frac{1}{2}}r\, \psi^{2}(\m{r}) +
{\frac{1}{4}}g\,\psi^{4}(\m{r})} \right]}} ,
\end{equation}
where all designations are standard and the field $\psi$ is
normalized so that the coefficient at the square
gradient is 1 \cite{book_patpokr,critbook_ma}. Such a form of the
fluctuational Hamiltonian is sufficient to describe the critical
asymptotical behavior of the free energy and is the basis for the
isomorphism of such behavior with that of the Ising model. Nevertheless,
the Hamiltonian lacks the asymmetrical terms which in general appear in the
description of the molecular fluids \cite{crit_hubbardschofield_pla1972,book_yukhgol_en}.

The asymmetry of the coexistence curve (CC) is the
characteristic feature for one-component liquids and solutions.
It is absent in the Ising model and numerous lattice models.
This circumstance indicates that the main role in the
appearance of the asymmetry belongs to the effects of finite
size for molecules of liquids.

The simplest characteristics of the asymmetry degree for the CC
is its diameter defined as
\begin{equation}
\label{diam_phi}
\varphi _{d} = {\frac{{1}}{{2}}}(\varphi _{l} +
\varphi _{\upsilon}  )\,,
\end{equation}
where $\varphi $ is the order parameter and subscripts $l,g$ stand for the coexisting
liquid and gas phases correspondingly. The latter can be identified
with the extensive variable suitable to differentiate between the coexisitng
liquid and vapor phases. As rule, the main
attention in literature is paid to the diameter of the CC in
terms density-temperature. Its investigation has a long
history.

In Pokrovskiy's work
\cite{crit_diamsingpokrovsky_jetplett1973} (see also \cite{crit_damsing_jetp1975})
it was shown that the classical law of the rectilinear diameter
\cite{crit_diam0,pcs_guggenheim_jcp1945} is violated:
\begin{equation}
\label{nd_1alpha}
\tilde{n}_{d} = \f{n_{l}+n_{g}}{2n_c} - 1 = D_{1 - \alpha}  \vert \tau \vert ^{1 - \alpha}  + D_{1}
\vert \tau \vert + \ldots\,,
\end{equation}
where $n$ is the number density, $\alpha $ is the critical exponent for the heat capacity.
The new term in Eq.~\eqref{nd_1alpha} is caused by the fluctuation
effects. In accordance with \cite{book_patpokr} this
contribution is connected with the non-orthogonality of the
fluctuations of the density and the energy. From another point
of view it means that the expansion of the density fluctuations
with respect to the algebra of strongly fluctuating variables
takes the form (see \cite{book_patpokr}):

\begin{equation}
\label{algebra_expansion}
\tilde {n} = \hat {A}_{1} + \lambda \hat {A}_{2} + \ldots\,
\end{equation}
Since $ \left\langle\,  \hat {A}_{1} \,\right\rangle = \pm
\,a_{1} \vert \tau \vert ^{\beta} $ (different signs correspond
to the liquid and vapor branches of the CC) and $
\left\langle\,  \hat {A}_{2} \,\right\rangle = a_{2} \vert \tau
\vert ^{1 - \alpha} $, the expansion Eq.~\eqref{algebra_expansion}
directly to Eq.~\eqref{nd_1alpha}.
%Note, that the operators $\hat {A}_{1}
%$ and $\hat {A}_{2} $ of strongly fluctuating variables are
%associated with the normal products of the order parameter
%$\psi $ for the Ising model:
%%
%\begin{equation}
%\label{algebra}
%\hat {A}_{1} ,\hat {A}_{2} ,\ldots\, \leftrightarrow
%:\psi :,\,:\psi ^{2}:,\ldots\,
%\end{equation}
%
For laboratory order parameter, which in general
is nonsymmetrical, the experimental
results should be fitted by the combination
\cite{crit_buckham_ptcp1972,crit_vdwscaling_physica1974}:
\begin{equation}
\label{ndiam_sing_2beta}
\tilde {\varphi}_{d} = D_{2\beta}
 \vert \tau \vert ^{2\beta}  + D_{1 - \alpha}
\vert \tau \vert ^{1 - \alpha}  +
D_{1} \vert \tau \vert + \ldots\,,
\end{equation}
where $\beta $ is the critical exponent for the CC in terms
temperature-density. The appearance of such singularity in
the diameter is trivial if $\varphi$
is a nonlinear function of some initial (laboratory) order parameter.
The appearance of the $\tau^{2\beta}$-singularity for the number density
is partly surprising. The density is the statistical average of
the simple microscopic field of the microscopic density and is
nothing but the unary distribution function. This means that
the density can be considered as the simplest irreducible
quantity in case of the liquid-vapor CP. So if the $2\beta$
anomaly exists for this quantity then the question about the
nature of $\tau^{2\beta}$-singularity of the density is deeper
and does not reduce to trivial algebraic relations between the
thermodynamic averages.

The revival of interest to the problem of $\vert
\tau \vert ^{2\beta} $- term is connected with the complete
scaling proposed in \cite{crit_fishmixdiam1_pre2003}. This approach
is based on the supposition that all three thermodynamic fields
temperature $\tau$, the chemical potential
$\tilde{\mu} = (\mu -\mu_c)/\mu_c$  and the pressure
$\tilde{p} = (P-P_c)/P_c $ are mixed equally.
Then it is easy to show that $\tilde {n}$ is a nonlinear
function of the average of the scaling fields which represent two
strongly fluctuating quantities. As a result,
the appearance of
the $\vert \tau \vert ^{2\beta} $- term in Eq.~\eqref{ndiam_sing_2beta}
is caused by the nonlinear algebraic dependence between $\tilde{n}$ and the average
of the order parameter. Numerous application of this approach are given in
\cite{crit_aniswangasymmetry_pre2007,crit_diamanis_cpl2006,
crit_asymsolut_jcp2010,crit_tolman_anisimov_prl2010}.
In particular, in complete scaling
approach the order parameter for the molecular liquids is the
nonlinear combination of the density field and the density of the entropy
 \cite{crit_aniswangasym_prl2006}. The density of the entropy is weekly
fluctuating field but the density field fluctuates more strongly.

%It is easy to see that the problem of the diameter singularity does
%not reduce to trivial algebraic dependence between the thermodynamic
%averages of the laboratory and the scaling order parameters. The existence
%of two strongly fluctuating fields the energy and the number density leads
%to the fact that  two order parameters fields exist - the specific entropy $s$ and
%the number density. They are nontrivially mixed.
%In the symmetrical case of the Ising model where these fields
%are orthogonal the corresponding diameters are:
%\begin{equation}\label{diam_sn}
%  \tilde{s}_d = A_{1-\alpha}\,|\tau|^{1-\alpha}+A_1\,|\tau|+\ldots \,,\quad  \tilde{n}_d = 0
%\end{equation}
%For asymmetrical case of liquids the diameters for these quantities
%have the same behavior given by Eq.~\eqref{ndiam_sing_2beta}
%\cite{crit_rehrmermin_pra1973}.
%%\begin{align}\label{diam_sn}
%%  \tilde{s}_d =& A_{2\beta}\,|\tau|^{2\beta} +
%%A_{1-\alpha}\,|\tau|^{1-\alpha}+A_1\,|\tau|+\ldots \,,\\
%%\tilde{n}_d =& D_{2\beta}\,|\tau|^{2\beta} +
%%D_{1- \alpha}\,|\tau|^{1-\alpha} + D_1\,|\tau| +\ldots
%%\end{align}

From another point of view the problem of the
singular terms in Eq.~\eqref{ndiam_sing_2beta} is solved in
\cite{crit_can_kul_cmphukr1997,crit_can_diamsing_kulimalo_physa2009}
on the basis of so called the canonical scaling. The last is
the synthesis of the scaling approach and the canonical
formalism, developed in the catastrophe theory
\cite{book_postonstewart}. In these works the regular procedure for
the construction of the fluctuation Hamiltonian for liquids,
which has the Ising-like form, is developed. Note, that the
Ising-like Hamiltonian coincides with that given by the standard
canonical form (see \cite{crit_can_kulinskii_jmolliq2003}).
As will be shown below
it corresponds to the thermodynamic potential with the only
one vanishing eigenvalue
of the stability matrix for the Gibbs potential.
The CC for liquids and solutions in the canonical variables is
fully symmetrical. One can say that the canonical order
parameter can be considered as the optimal one. The
asymmetry of the CC appears only if one returns to the
laboratory variables. The evident relations between the
canonical and laboratory order parameters as well as between
the canonical and laboratory fields are established. The
distinctive peculiarity of these relations is their nonlinear
character. Nevertheless, in contrast to the complete scaling approach the canonical one
operates with two fluctuating fields and uses the nonlinear transformation
rather for the field variable of the order parameter than the
nonlinear transformation for its average
value \cite{crit_can_diamsing_kulimalo_physa2009}.

The motivation of this work is the search for the unifying
Hamiltonian formulation
of the problem of the asymmetry suitable
for the determination of the proper order parameter in case of two
fluctuating fields one of which fluctuates less than the other.

\section{Gibbs potential analysis}
Now we discuss briefly how the order parameter
$\varphi$ and the field $h$ conjugated to
it, as well as the value of $r$ in Eq.~\eqref{lg_ham}
 are connected
with the expansion of the fluctuational deviation
for the Gibbs potential $g(n,s)$ near the critical point:
\begin{equation}
\label{gibs_zetaphi}
\delta g = {\frac{{1}}{{2}}}a_{2,0} \varphi ^{2} +
{\frac{{1}}{{2}}}a_{0,2} \zeta ^{2} +
{\frac{{1}}{{3}}}a_{3,0} \varphi ^{3} +
{\frac{{1}}{{3}}}a_{2,1} \varphi ^{2}\zeta  +
{\frac{{1}}{{3}}}a_{1,2} \varphi \zeta ^{2} +
{\frac{{1}}{{3}}}a_{0,3} \zeta ^{3} +
{\frac{{1}}{{3}}}a_{4,0} \varphi ^{4} + ...
\end{equation}
Here
\begin{align}
a_{2,0} =& \left. {{\frac{{\partial \mu}} {{\partial n}}}} \right|_{T}
- \f{\left(\,\left.\partial T / \partial n\right|_{s}\,\right)^2}
{\left.\partial T / \partial s\right|_{n}}\,,\quad
a_{0,2} = \left.
\frac{\partial\, T}{\partial\, s}\right|_{n}=
{\frac{{T}}{{c_{\upsilon}}}}
\end{align}
\[\varphi  = n - n_{c} \,,\quad  \zeta  =
s - s_{c} +
\f{c_v}{T}\left.\partial T / \partial n\right|_{s}(n - n_{c} )\,,\]
where
$\mu $ is the chemical potential, $c_{\upsilon}  $
is the heat capacity, $n$ and $s$ are the density and entropy
correspondingly, the subscript ``$c$``
denotes their values at the critical point.

Usually it is taken into account that only one of two
coefficients $a_{2,0}$ and $a_{0,2}$ in \eqref{gibs_zetaphi}
vanishes at the critical point. The supposition
\begin{equation}\label{a11}
a_{2,0} = \left. {{\frac{{\partial \mu}} {{\partial n}}}} \right|_{T} \to
0\,,\quad T \to T_{c}\,\,.
\end{equation}
corresponds to the situation when only the fluctuations
of the density become anomalously large.
Note that in the framework of the thermodynamic fluctuation
theory the second coefficient $a_{0,2}\ne 0 $. Indeed, in
the vapor phase
$c_{\upsilon}  $ takes the finite value and the transition to
liquid state is accompanied by the addition of the only degree of
freedom (the order parameter), which leads to the increase of
$c_{\upsilon}  $ on ${\frac{{1}}{{2}}}k_{B}$.

Since $a_{0,2}\ne 0 $, the fluctuations of $\zeta $ are
bounded and we can neglect them thus reducing \eqref{gibs_zetaphi}
to the standard Landau expansion:
\begin{equation}
\label{gibbspotential}
\delta g_{L} = {\frac{{1}}{{2}}}a_{2,0} \varphi ^{2} +
{\frac{{1}}{{3}}}a_{3,0} \varphi ^{3} + {\frac{{1}}{{4}}}a_{4,0}
\varphi ^{4} + {\frac{{1}}{{5}}}a_{5,0} \varphi ^{5} + \ldots\,\,.
\end{equation}

In fact the local part of Eq.~\eqref{lg_ham}
is nothing but the local form of Eq.~\eqref{gibbspotential}
truncated to the minimal polynom sufficient to describe
the reconstruction of the minima which represent the
coexisting phases.

The transformation from the infinite series \eqref{gibbspotential}
to the canonical form \eqref{lg_ham} is discussed in details
in \cite{crit_can_kulinskii_jmolliq2003,
crit_can_diamsing_kulimalo_physa2009}. In that case when $c_v$
is bounded, the canonical form for the local part of the
fluctuation Hamiltonian takes the form:
\begin{equation}\label{a3cat}
h_{loc}(\psi) = \left.\delta g_{L}\right|_{\varphi \to
\varphi(\psi)} = \f{1}{4}\,g\,\psi^4+\f{1}{2}\,r\,\psi^2+
h_{\psi}\,\psi\,\,.
\end{equation}
The canonical order parameter $\psi $ is the nonlinear
function of initial variable $\varphi$:
\[\psi = \varphi + \f{1}{2}\,\gamma_2\varphi^2+\f{1}{2}\,
\gamma_3\varphi^3+ \ldots\,\,.\]
Explicit expressions for $g,r$ and $h_{\psi}$ as well as
for the coefficients $\gamma_n$c an be found in \cite{book_gilmore1}.

%The core element here is the notion of the order parameter which is the anomalously fluctuating quantity so that the thermodynamic quasipotential depend on the only ``density`` variable $\varphi$ \cite{crit_griffitswheeler_pra1970}. All
%other densities, e.g. entropy are assumed to depend on the
%order parameter since their fluctuations are bounded.
%Once the number of the anomalous fluctuating densities is known the derivation of the Landau quasipotential can be performed in the framework of the catastrophe theory
%\cite{book_postuart,book_gilmore1} where the canonical forms
%for the potential functions are given. In particular the Landau minimal $\varphi^4$-potential coincides with the canonical form of $A_3$ cusp catastrophe \cite{book_gilmore1}:
%
%
%The mean-field description (Landau theory) neglects
%the fluctuations of the second field thus leading
%to the jump of the heat capacity which allows to
%differentiate between phases at the critical point.
%Sure this property is in conflict with the nature
%of the critical state where the coexistence between
%phases terminates and they become indistinguishable.

However, in many cases the applicability region of the
thermodynamic fluctuation theory or the mean-field
approximation is absent. Argon can serve as the simplest
example of such a kind. It is evident from the behavior
of its isothermal compressibility. In other words the
Ginsburg temperature for argon and many other liquids is
close to unity $\tau_{Gi} (Ar)\sim (0.1\div 0.3)$
\cite{book_anisimov}. It means that the behavior of the heat
capacity $c_{\upsilon}  $ is determined by fluctuation effects.
Therefore, fluctuation effects cannot be ignored and they
lead to the vanishing of the quantity $T/c_v$, which is
the second eigenvalue of the quadratic form in Eq.~\eqref{gibs_zetaphi}.
In such a case the canonical form for the local part of the
fluctuation Hamiltonian should be generalized since both
variables $\varphi $ and $\zeta $ should be considered as the
strongly fluctuating ones.

The new version of the fluctuational Hamiltonian
will be able to reflect more exactly the relative role of
mechanical and thermal variables, which are used for the
description of the critical behavior. At that, if an order
parameter is modeled by a mechanical variable, the auxiliary
field in the fluctuational Hamiltonian is given by caloric
one and vice versa. In particular, the specific volume as the
order parameter is completed by the specific entropy. Taking
into account of this variable is especially important for complex
liquids, molecules of which rotate, can form H-bonds and have internal
degrees of freedom excited in the thermal motion, as well as for
multicomponent solutions. Since fluctuations of the specific volume
and specific entropy are not orthogonal, the increase of the role of
caloric effects should be accompanied by a growth of the coexistence
curve asymmetry.
In connection with the last problem, it is necessary to note that the
asymmetry of coexistence curves arises not only due to caloric effects.
Another mechanism is caused by hard core effects. Their manifestation in
complex liquids is similar to that for simple liquids.
Therefore in this paper our
attention will be mainly focused on the inclusion of the
caloric effects into the fluctuational Hamiltonian.

Therefore the fluctuational Hamiltonian should be built
on the basis of the canonical form for the thermodynamic
potential which is consistent with the fact of the existence
of two strongly fluctuating fields.
Another support for such conjecture is given by the two-dimensional
lattice models where the requirement of the conformal symmetry
leads to the two independent strongly
fluctuating fields
\cite{cft_bpz_nuclphys1984,cft_difranchesko_mathieu_senecal}.
From here and the very general isomorphism principle it
follows that the bare fluctuation Hamiltonian for liquids
should also contain two independent order parameters.

\section{Fluctuation Hamiltonian of the system}\label{sec_fluctham}
The local part of the fluctuation Hamiltonian for a
system, for which two eigenvalues $a_{2,0}$ and $a_{0,2}$ of
the stability matrix tend to zero at the critical point, in
accordance with the
catastrophe theory must be built on the basis of $D_5$
catastrophe \cite{book_gilmore1}:
\begin{equation}\label{d5cat}
\left.\delta g\right|_{\varphi\to\psi(\varphi,\zeta),\zeta\to
\sigma(\varphi,\zeta)} = \f{1}{4}\,\psi^4 +b\,\psi\,\sigma^2 +
\f{1}{2}\,r\,\psi^2+ \f{1}{2}\,u\,\sigma^2 +h_{\psi}\,\psi +
h_{\sigma}\,\sigma\,,
\end{equation}
where $\psi$ and $\sigma $ are the corresponding canonical
variables which are the nonlinear functions of the initial
variables $\varphi$ and $\zeta$. In view of the results of
Ref.~\cite{crit_can_diamsing_kulimalo_physa2009}
the parameter of the asymmetry coupling
$b$ canonically represents the degree of the
asymmetry of the initial Hamiltonian.

The fluctuational Hamiltonian consists of two parts:
\[\mathcal{H} = H_{loc} + H_{qloc}\,,\]
the local part $ H_{loc}$
\[H_{loc} = \int \delta g(\psi(\m{r}),\sigma(\m{r}))\,d\m{r}\]
and the quasilocal part $ H_{qloc}$. For the latter we use
common quadratic gradient approximation.
In accordance with Eq.~\eqref{d5cat} the local part of the
fluctuational Hamiltonian %
has the form:
\begin{equation}
\label{ham_d5}
H_{loc} [\psi (\m{r}),\sigma  (\m{r})] =
H_{loc} [\psi  (\m{r})] +
H_{loc} [\sigma (\m{r})] +
H_{int} [\psi  (\m{r}),\sigma (\m{r})]\,\,,
\end{equation}
with
\begin{align}
H_{loc} [\psi  (\m{r})] =& \int d\m{r}\left[ { - h_{\psi }\,  \psi +
\frac{{1}}{2}r\, \psi^{2} + \frac{{1}}{4}g\,\psi^{4}} \right]\,,
\label{ham_d5_phi}\\
H_{loc} [\sigma (\m{r})] = & \int {d\m{r}{\left[ { - h_{\sigma}\,
 \sigma + {\frac{{1}}{{2}}}u\, \sigma^{2}} \right]}} \,,
 \label{ham_d5_zeta}\\
H_{int} [\psi (\m{r}),\sigma (\m{r})] =&
\int {d\m{r}\,b\,\psi\, \sigma^{2}}\, \label{ham_d5_zetaphi}.
\end{align}
Here $\psi(\m{r})$ and $\sigma(\m{r})$ are the fluctuating
fields. In accordance with Eq.~\eqref{gibs_zetaphi}
$\psi(\m{r})$ can be identified with the density
while $\sigma(\m{r})$ is the combination of the
density and the entropy fluctuations. The local part
of the Hamiltonian should be completed with the relevant
gradient terms. They are
\begin{gather}
H_{nloc}[\psi(\m{r})] = \int \f{1}{2}\,
\left(\,\nabla \psi (\m{r})\,\right)^2 \, d\m{r}\,\,, \label{hqloc1}\\
H_{nloc}[\psi(\m{r})] = \int
\f{c_{\psi\,\sigma }}{2}\,\nabla \psi(\m{r})\cdot \nabla \sigma (\m{r})\, d\m{r}\,\,.\label{hqloc2}
\end{gather}
We did not include the quadratic term
$\left(\nabla \sigma \right)^2$ because
as we show below within the dimensional
analysis it is irrelevant.

\subsection{Simple dimensional analysis}\label{subsec_dim}
Here we perform the analysis of the Hamiltonian
\eqref{ham_d5} introduced
above using the simple scaling \cite{book_scaling_cardy}.
From the elementary analysis of dimensions and the fact that the
main contribution to the
coefficient $u$  are caused by the fluctuations of the free
field $\varphi$ it follows that its dimension is
$\Delta^{(0)}_{u} = \Delta^{(0)}_{g} = 4-d= \epsilon$. The character of
the temperature dependence of $r$ does not change
in comparison with \eqref{lg_ham}, so
$\Delta^{(0)}_{r} = 2$ and $r = r_d\,\tau $.
Also it is obvious that $\Delta^{(0)}_{\sigma} =
2\Delta^{(0)}_{\psi} = d-2$. Thus for the
conjugated field
$h_{\sigma}$ we have
$\Delta^{(0)}_{h_{\sigma}} = d- \Delta^{(0)}_{\sigma}= 2$,
thus it is nothing but the canonical temperature
variable. The field $h_{\psi}$ obviously plays the
role of the canonical chemical potential and the condition
$h_{\psi} = 0$ determines the coexistence line.
The scaling dimension of $b$ - the strength of
asymmetric interaction is $\Delta^{(0)}_{b} =
\f{3}{2} \left(\, \f{10}{3}  - d\,\right) $,
which becomes relevant below $d=\f{10}{3}$ and
thus equivalent to the quintic $\psi^5$
asymmetrical term in the standard consideration
\cite{crit_nicollzia_prb1981}.
The gradient term in Eq.~\eqref{hqloc2} has the same
dimension as the
asymmetric term $\varphi\,(\nabla \psi)^2$,
which should be taken into account along with
the quintic coupling \cite{crit_nicoll_pra1981}.
The term $(\nabla \sigma)^2$ can be omitted
because it is irrelevant
(with dimension
$\Delta^{(0)}_{\nabla \sigma} = 2-d = -2+\epsilon $).

For the Ising model because of the
symmetry $\psi \to -\psi$, $b = c_{\psi\,\sigma} = 0$
and the fields $\psi$ and $\sigma$ decouple.

Previous consideration of the classical scaling dimensions
obviously allows to identify $\sigma$ with the field
$:\psi^2(\m{x}):$. With account of the fluctuations the
dimensions are renormalized
$\Delta_{\psi} = \Delta^{(0)}_{\psi}+\f{\eta}{2}$ and
$\Delta_{\sigma} = \Delta^{(0)}_{\sigma} +2-\f{1}{\nu}$
so that $\Delta_{h_{\sigma}} = 1/\nu $ and indeed
$h_{\zeta}$ is a temperature field. Here $\nu$ is the critical
exponent of the correlation length and $\eta$ is the Fisher's
critical exponent of the anomalous dimension
\cite{crit_wilsonfisher_prl1972}. The dimension
of the parameter $u$ is $\Delta_{u} = \f{2}{\nu} - d =
\f{\alpha}{\nu}$. This leads to the conclusion that
indeed $u \propto \tau^\alpha$, i.e. asymptotically it
is the inverse value of the specific heat.

\section{Fluctuation Hamiltonian for binary solution}
\label{sec_binarymix}

In this Section we briefly discuss the specificity of the fluctuation
Hamiltonian in binary solutions. The method of consideration is naturally
generalized for arbitrary multi-component solutions.

Let us consider the binary solution, for which the total
number of particles
remains to be constant. In this case the thermodynamic
state of the system
is set by the temperature and pressure as well as the
difference of the
chemical potentials of components:
$\mu = \mu _{2} - \mu _{1} $. Considering
$\mu $ as a function of specific volume $\upsilon $, entropy and
concentration $x$ we can write the following
expression for the increment of
the chemical potential near its minimal value:

\begin{equation}
\label{eq1}
\delta ^{2}\mu =
{\frac{{1}}{{2}}}{\left[ {{\frac{{M_{3}}} {{M_{2}}} }\delta
x^{2} +
{\frac{{M_{2}}} {{M_{1}}} }(\delta x + c\delta \upsilon )^{2} +
M_{1} (\delta x + a\delta s + b\delta \upsilon )^{2}} \right]} + ...
\end{equation}

Here $M_{i} ,\,\,\,i = 1,2,3$, are the main minors of the matrix:

\[
\hat {M} = \left( {{\begin{array}{*{20}c}
 {\partial T / \partial s} \hfill & {\partial T / \partial \upsilon}  \hfill
& {\partial T / \partial x} \hfill \\
 {\partial P / \partial s} \hfill & {\partial P / \partial \upsilon}  \hfill
& {\partial P / \partial x} \hfill \\
 {\partial \mu / \partial s} \hfill & {\partial \mu / \partial \upsilon}
\hfill & {\partial \mu / \partial x} \hfill \\
\end{array}}}  \right).
\]

In fact, the diagonal form (\ref{eq1}) is naturally
obtained by the Lagrange method.

The coefficients $a,\,b$ and $c$ are some
combinations of the elements of
$\hat {M}$.

It is not difficult to verify that near the mixing critical point:
\[{\frac{{M_{3}}} {{M_{2}}} }
\sim {\left. {{\frac{{\partial \mu}} {{\partial
x}}}} \right|}_{P,T}\,\quad \text{and} \quad
{\frac{{M_{2}}} {{M_{1}}} }\sim
{\frac{{T}}{{c_{\mu \upsilon}} } }\]

Since $M_{1} = {\frac{{T}}{{c_{x\upsilon}} } }$ and
$c_{x\upsilon}  $ takes
the finite value for binary solution the fluctuations
of the combination
$\delta u = \delta x + a\delta s + b\delta \upsilon $
are bounded and the
contribution of corresponding term in Eq.~\eqref{eq1}
can be omitted.

Let us compare (\ref{eq1}) with the analogous increment
of the chemical potential
for one-component liquid near its vapor-liquid critical point.
In this case
\[
\delta ^{2}\mu = {\frac{{1}}{{2}}}{\left[ {{\frac{{M_{2}}} {{M_{1}}} }\delta
\upsilon ^{2} + M_{1} (\delta s + a\delta \upsilon )^{2}} \right]} + ...{\rm
,}
\]
where
\[
\hat {M} = \left( {{\begin{array}{*{20}c}
 {\partial T / \partial s} \hfill & {\partial T / \partial \upsilon}  \hfill
\\
 {\partial P / \partial s} \hfill & {\partial P / \partial \upsilon}  \hfill
\\
\end{array}}}  \right),
\]
and
${\frac{{M_{2}}} {{M_{1}}} } = - {\left. {{\frac{{\partial P}}{{\partial
\upsilon}} }} \right|}_{T} $ and $M_{1} = {\frac{{T}}{{c_{\upsilon}} } }$.

As we see, the characters of the near-critical behavior of $\delta ^{2}\mu $
for binary solutions and one component liquids are fully identical.
Therefore we conclude that critical phenomena in them are isomorphic, i.e.
their fluctuation Hamiltonians have the same form.

\section{Conclusions}
We have performed the stability analysis of the
thermodynamic potential for
the molecular systems with the critical point
where $\tau_{Gi}$ is not small $\tau_{Gi}\gtrsim 0.1$.
Such situation means that there is no crossover region
to the mean-field behavior so the canonical form for the fluctuational
Hamiltonian include two independent fluctuating fields.
Both fields are strongly fluctuating but one of them
has the strongest singularity caused by the slowest
decay of its correlator. The proposed canonical form
of the fluctuational Hamiltonian is constructed on the basis of
$D_5$ catastrophe and contain asymmetrical term which
describes the
coupling between these fields. If the contribution of
less fluctuating field is neglected we get the standard
Landau-Ginzburg-Wilson Hamiltonian.
The facts that $\tau_{Gi}$ is not small and the leading
correction to the quadratic part (random phase approximation)
is determined by $\psi^4$ term allow to conclude about
global cubic character of the binodal. This is in
accordance with the global isomorphism proposed in
Ref.~\cite{eos_zenome_jphyschemb2010}.

Simple scaling analysis shows that one of the field
can be identified with the density-like one while
another field corresponds to the fluctuations of the
entropy. We hope that the proposed canonical form
can serve as the basic fluctuational Hamiltonian which
describes the asymmetry effects characteristic for
the real fluids more adequately. The problem of
derivation of \eqref{ham_d5} on the basis of microscopic
Hamiltonian can be performed within the canonical approach
\cite{crit_can_kulinskii_jmolliq2003,
crit_can_diamsing_kulimalo_physa2009}.

The physical reasonings for the appearance of two fluctuating fields
are connected with the account of the hard core effects in the
microscopic Hamiltonian. Usually the
microscopic Hamiltonian is divided into short-range and
long-range parts due to the corresponding division of the
interaction potential into repulsive and attractive ones.
The repulsive part is usually integrated over and the
fluctuational part depending only on the density fluctuations
appears (see e.g. Refs.~\cite{crit_hubbardschofield_pla1972,book_yukhgol_en}).
The asymmetry of the coexistence curves for real liquids arises
due to the breaking of the particle-hole symmetry for the lattice gas.
This symmetry is first of all violated because of different role
of the hard core effects in liquid and vapor phases.
It is also violated by different manifestation of caloric
effects in coexistence phases.

It is necessary to stress that the canonical fluctuation
Hamiltonian \eqref{ham_d5} of liquids is not
isomorphic to that for
the Lattice Gas or equivalently for the Ising model,
since the interaction Hamiltonian \eqref{ham_d5_zetaphi}
is an odd function of the order
parameter:
$H_{int} [ - \psi (\m{r}),\sigma (\m{r})] = - H_{int} [\psi
(\m{r}),\sigma (\m{r})]$. We can speak about
isomorphism between them
only if $H_{int} [\psi (\m{r}),\sigma (\m{r})]$
is negligibly small. The results of
Ref.~\cite{crit_can_diamsing_kulimalo_physa2009}
allow to connect the parameter of the asymmetry coupling
$b$ with the asymmetry of the initial Hamiltonian on the
basis of the model for the equation of state for the real
molecular fluids. With such model equation of state the
caloric and the hard core effects can be taken into account
explicitly. The detailed consideration of this question will be a
subject of prospective
work.%
\section*{Acknowledgements}
The authors thank Prof. M. Kozlovsky,
Prof. M.~Anisimov and Dr. C.~Bertrand
for discussion during PLMMP-2010, Kiev 23-26 June.

%\newpage
%\bibliography{criticality,conform,books_my,thermodyn}

\end{document}